\documentclass[12pt,preprint]{aastex}

\lefthead{Escala}
\righthead{Evidence in Favor of Direct Massive Black Hole Formation.}

\begin{document}

\title{Universal Relation for  Life-span  Energy Consumption in Living Organisms: 
Insights for the 
origin of ageing.}

\author{Andr\'es Escala}
\affil{Departamento de Astronom\'{\i}a, Universidad de Chile, Casilla 36-D, Santiago, Chile.}
\affil{aescala@das.uchile.cl}

\begin{abstract}


{\bf  Metabolic energy consumption has long been thought to play a major role in the aging process ({\it 1}). Across species, a gram of tissue on average expends about the same amount of energy during life-span  ({\it 2}). Energy restriction has also been  shown that increases maximum life-span ({\it 3}) and   retards age-associated changes ({\it 4}). However, there are significant exceptions to a universal energy consumption during life-span, mainly coming from the inter-class comparison ({\it 5, 6}). Here we present a unique  relation for  life-span  energy consumption, valid  for $\sim$300 species representing all classes of living organisms, from unicellular ones to the largest mammals. The relation has  an  average scatter of only 
 0.3 dex, with 95\% ($\rm 2-\sigma$) of the organisms having  departures
less than a  factor of $\pi$ from the relation, despite the $\sim$20 orders of magnitude difference in  body mass, reducing any possible  inter-class variation in the relation to only a geometrical factor.
This result can be interpreted as supporting evidence  for the existence of an approximately constant total number $\rm N_r \sim 10^8$ of respiration cycles per lifetime for all organisms, effectively  predetermining 
the  extension of life by the basic energetics of respiration, being  an 
 incentive for
 future  
 studies that investigate the  relation of such constant  $\rm N_r$ cycles per lifetime with the production rates of free radicals and oxidants, which may give definite constraints on the origin of ageing.}
\end{abstract}

\section{Introduction}

Ageing seems to be   an inherent characteristic of all living organisms, while some  inert  objects can easily subsist  on Earth  for many centuries,  living systems have a much narrow existence, typically limited to  decades for large animals. Is then somewhat natural  trying  to associate the process of ageing  to   metabolism, since   all living organisms  get the required energy  to stay alive from  such  process. Rubner in 1908 ({\it 7}),  compared  energy metabolism and lifespans of five domestic animals (guinea pig, cat, dog, cow and horse) and man, finding  that the life-span (total) energy expenditure per gram for the five species is approximately constant, suggesting  the total metabolic energy consumption per lifespan is fixed, which later has become known as the `rate of living'  theory ({\it 1}). 

 Decades later,  a  mechanism was found  in which the idea behind a fixed energy consumption per lifespan might work  in the `free-radical damage' 
hypothesis  of aging  ({\it 8, 9}), in which   macromolecular components of the cell are under perpetual attack from toxic by-products of metabolism, such as free radicals and oxidants. 
In addition, energy restriction has been experimentally shown to increase maximum life-span and retard age-associated changes in  animals, 
such as insects, rats, fish, spiders, water fleas and mice ({\it 3, 4}).

Rubner's relation was confirmed for around hundred mammals  ({\it 10}) and extended   it to  birds ({\it 11}), ectotherms ({\it 12}) and even unicellular organisms   such as protozoa and bacteria ({\it 6}), totalizing almost three hundred different species in a range of   20 orders of magnitude  in body mass. Although  the total metabolic energy exhausted per lifespan per  body mass of given organism  appears to be relatively constant parameter, at about the same number determined by Rubner ({\it 7}) of a million  
Joule per gram of body weight for mammals, variations over an order of magnitude are found among different animal classes, a result also found by other authors (i.e. {\it 2}) and  considered the most persuasive  evidence against the `rate of living'  theory ({\it 5}).

The origin of such  variations in the lifespan energy consumption   might come from intrinsic variations in the quantities used to estimate it: lifespans and especially,  the metabolic rate relation  which universality  is still debated (see {\it 13} and references therein). Recently, the empirical metabolic rate relation was corrected in order to fulfill dimensional homogeneity ({\it 13}), a minimal requirement of any meaningful law of nature  ({\it 14}), proposing a new metabolic rate (B) formula: $\rm B  =  \epsilon(T) \, \eta_{O_2} f_H  \,  M$, where M is the body mass,  $\rm f_H $ is a characteristic (heart) frequency, $\rm \eta_{O_2} $ is  an specific  $\rm {O_2} $ absorption factor and $\rm \epsilon(T)=\epsilon_0 \, {\large e^{\small -E_a/k T}}$ is a temperature correction inspired in the  Arrhenius formula, in which $\rm E_a$ is an activation energy and k is the Boltzmann universal constant. Compared to  Kleiber's original formulation ({\it 15}), $\rm B  = B_0 (M/M_0)^{0.75}$,  this new metabolic rate relation  has the  heart frequency  $\rm f_H $ as independent controlling variable (
a marker of metabolic rate) and the advantage of being an unique  metabolic rate equation  for different classes of animals  and different exercising  conditions, valid for both basal and maximal metabolic rates, in agreement with  empirical  data in the literature ({\it 13}).

In addition, this new metabolic rate relation can be directly linked  to the total energy consumed in a lifespan ({\it 13}), being a promising way to explain the origin of variations in Rubner's relation and unify  them into a single formulation. In this paper, we will explore  the implications of this new metabolic relation for the total energy consumed in a lifespan and the `rate of living'  theory,  being organized as follows. We start reviewing  the results  of the new metabolic rate relation found in ({\it 13}),  deriving  its prediction for the total energy consumed in a lifespan in \S2. Section 3  continues with  testing the empirical support of the predicted relation  for total energy consumed in a lifespan, with satisfactory results. Finally in \S4, we discuss the results and implications of this work.


\section{New Metabolic Rate Relation and its Prediction for `Rate of Living'  Theory}


In the proper mathematical formulation of natural laws, a {\it minimum} requirement is to be expressed in a general form that remain true when the size of units is changed, simply because nature cannot fundamentally depend on a human construct such as  the definition of units. In mathematical terms, this implies that meaningful laws of nature must be homogeneous equations in their various units of measurement  ({\it 14}). However, most relations in allometry do not satisfy this basic requirement, including Kleiber's `3/4 Law' ({\it 15}), that relates the basal metabolic rate and body mass as $\rm B  = B_0 \, (M/M_0)^{3/4} 
= C \, M^{3/4}$, being this  `3/4 Law'  a typical  example ({\it 14}) of a relation where  the proportionality factor C has a fractal dimensionality  
and its value depends on the units chosen for the variables ($\rm B,  M$), therefore, do not fulfill  the   {\it minimum}   requirement for being a  natural law
. Mathematically speaking,  to qualify for being a natural law the metabolic rate relation must be first rewritten such the constants with dimensions are  universal and restricted to a minimum number, which in no case could exceed the total number of fundamental units of the problem.

To solve this issue, ({\it 13}) proposed a new unique homogeneous equation for the metabolic rates that includes the heart frequency  $\rm f_H $ as independent controlling variable, proposing a the metabolic rate (B) formula: $\rm B  =  \epsilon(T) \, \eta_{O_2} f_H  \,  M$, where M is the body mass,  $\rm f_H $ is a characteristic (heart) frequency, $\rm \eta_{O_2} $ is  an specific  $\rm {O_2} $ absorption factor and $\rm \epsilon(T)$ is a temperature-dependent normalization. The new metabolic rate relation is, in addition to be  in agreement with the empirical data  ({\it 13}),    valid for different classes of animals and for both resting and exercising conditions. Using this formula, it can be shown   ({\it 13})
    that most of the differences found in the allometric exponents  are due to compare incommensurable quantities, because the variations in the dependence of the metabolic rates on body mass are secondary, coming from variations in the allometric dependence of the heart frequencies   $\rm f_H $ on the body mass M. Therefore,  $\rm f_H $ can be seen as a new independent physiological variable that
controls metabolic rates, in addition to the body mass.

One of the advantages of having a metabolic rate relation with  the heart frequency  $\rm f_H $ as independent controlling variable,  is that it can be straightforwardly  linked ({\it 13}) to the total energy consumed in a lifespan, by the relation of total number $\rm N_b$ of heartbeats in a lifetime, $\rm N_b =   f_H \,   t_{life}$, which is empirically determined to be constant for mammals and equals to $\rm 7.3 \times 10^8$ heartbeats ({\it 16})
.  Therefore, for a constant total number  of heartbeats in a lifetime, $\rm t_{life}  =   N_b \,  / \, f_H$, the  metabolic rate relation ({\it 13}) can be rewritten as 
$\rm B    \,   t_{life} =  \epsilon(T) \, \eta_{O_2} N_b \,  M $, giving a straightforward prediction for  life-span energy consumption and that can be seen as a   test for the `rate of living' theory ({\it 1}).

However, since the  total energy consumed in a lifespan relation is valid even in unicellular organisms without heart ({\it 6}), in order to find an unique relation for all living organisms the concept  of characteristic (heart) frequency must be first  generalized. A natural candidate is   the  respiration frequency, $\rm f_{resp}$, since this frequency is observed in animals to be strictly proportional to the heart one, $\rm  f_H = a \, f_{resp}$ ({\it 17}), and is a still meaningful frequency for organisms without heart. Under this proportionally between frequencies,   the empirical relation with lifetime can be rewritten to be  also  valid for a total number $\rm N_r \,(=N_b/a)$ of `respiration cycles': $\rm  t_{life} = N_b/f_H = N_b/af_{resp} = N_r/f_{resp}$.  This total number of `respiration cycles', $\rm N_r = f_{resp} t_{life} $,  will be assumed  from now to be the same number for all living organisms  and  in \S 4, we will   explore the implications of this conjecture for the origin of aging.

Under the 
condition of a constant total number $\rm N_r $ 
of respiration cycles in a lifetime $\rm  t_{life}$, 
multiplying  the new metabolic rate relation ({\it 13}) by $\rm  t_{life} /a = N_r/af_{resp} = N_r/f_{H}$ is now 
equivalent to B$\rm \,t_{life}/a = \epsilon(T) \, \eta_{O_2} N_r  \,  M$.
The factor $\rm  \epsilon(T) \, \eta_{O_2} =  \epsilon_0 \, \eta_{O_2} \, {\large e^{\small -E_a/k T}} $ can be rewritten as   $\rm  E_{2019} \,    {\large e^{\small (\frac{1}{T_a}-\frac{1}{T})\frac{E_a}{k } }} 
$, where $\rm T_a$ in a normalizing `ambient' temperature  and  $\rm E_{2019}= 10^{-4.313} \,mlO_2g^{-1} 
\approx  10^{-3} \, Jg^{-1} $ (converting  1 ltr $\rm O_2$=20.1 kJ; {\it 17}) a constant  that comes from  the best fitted value for the corrected metabolic relation ({\it 13}). Therefore, for the metabolic rate relation given in ({\it 13}) and under the condition of fixed respiration cycles in a lifetime, the following relation is predicted to   be valid:

\begin{equation}
\rm  
exp\Big({\small \frac{E_a}{k T}}\Big) \,  \frac{  {B \, t_{life}}  }{a} 
 =  E_{2019} \,   exp\Big({\small \frac{E_a}{k T_a}}\Big) \, N_r  \,  M \, .
\label{Eq1}
\end{equation}



Eq. \ref{Eq1} is a prediction from the mathematically-corrected metabolic relation ({\it 13}) under the assumption of constant  $\rm N_r $ respiration cycles in a lifetime, $\rm  t_{life} = N_r/f_{resp}$, which is in principle valid for all living organisms. It is important to emphasize that this relation for lifespan energy consumption is not assumed or hypothesized to be fixed like in the `rate of living' theory, instead is  derived directly from  the metabolic relation under the conjectured invariant $\rm N_r $. Compared to allometric relations, characterized mathematically for having as many dimensional constants ($\rm B_0,  M_0$, in Kleiber's case) as there are variables ($\rm B,  M$), the relation given by Eq. \ref{Eq1}  has (in principle) 7 variables $(\rm  t_{life}, B, a, M, E_a, T_a \, \& \, T)$, one dimensionless number  ($\rm N_r $) and  only two  constants with units: one universal one (Boltzmann  constant k)  and another one    coming from best fitting  the  energetics of respiration $(\rm E_{2019})$, which possible universality will be discussed in \S4.

\section{Empirical Support for the Corrected Relation of Total Energy Consumed in a Lifespan}

In this section we test the validity and accuracy of the derived relation for total energy consumed in a lifespan (Eq. \ref{Eq1}), predicted from the new formulation of the metabolic rate relation ({\it 13}) and the conjectured constant total number $\rm N_r$ of respiration cycles per lifespan for all living organisms. For that purpose, we use data from 
277 species of all classes of  living organisms, from unicellular organisms and other ectotherms species, to mammals and birds,  listed in Table 1 of ({\it 6}) with their  body mass M, total metabolic energy per lifespan $\rm B \,  t_{life}$ and body temperature T. 

Fig 1 shows the  relation predicted by Eq. \ref{Eq1} for the 277 living organisms   listed in  ({\it 6}), where the  activation energy $\rm E_a$ was chosen  to the average  value of 0.63 eV, independently   determined to temperature-normalize the metabolic rates of  unicells and poikilotherms   to endotherms ({\it 18, 19}). The parameter  a was chosen to the empirically determined values of 4.5 for  mammals and 9 for birds ({\it 17}), estimated to be  3 for ectotherms with heart (from the relative size of their hearts; {\it 17}) and assumed unity  for ectotherms without heart, such as unicellular organisms. The total number $\rm N_r =N_b /a = 1.62 \times 10^8$ of respiration cycles in a lifetime, was determined from the best fitted values for mammals: $\rm N_b=7.3 \times 10^8$ heartbeats  in a lifetime ({\it 16}) and a=4.5 ({\it 17}). 

\begin{figure}[h!]
\begin{center}
\includegraphics[width=11.9cm]{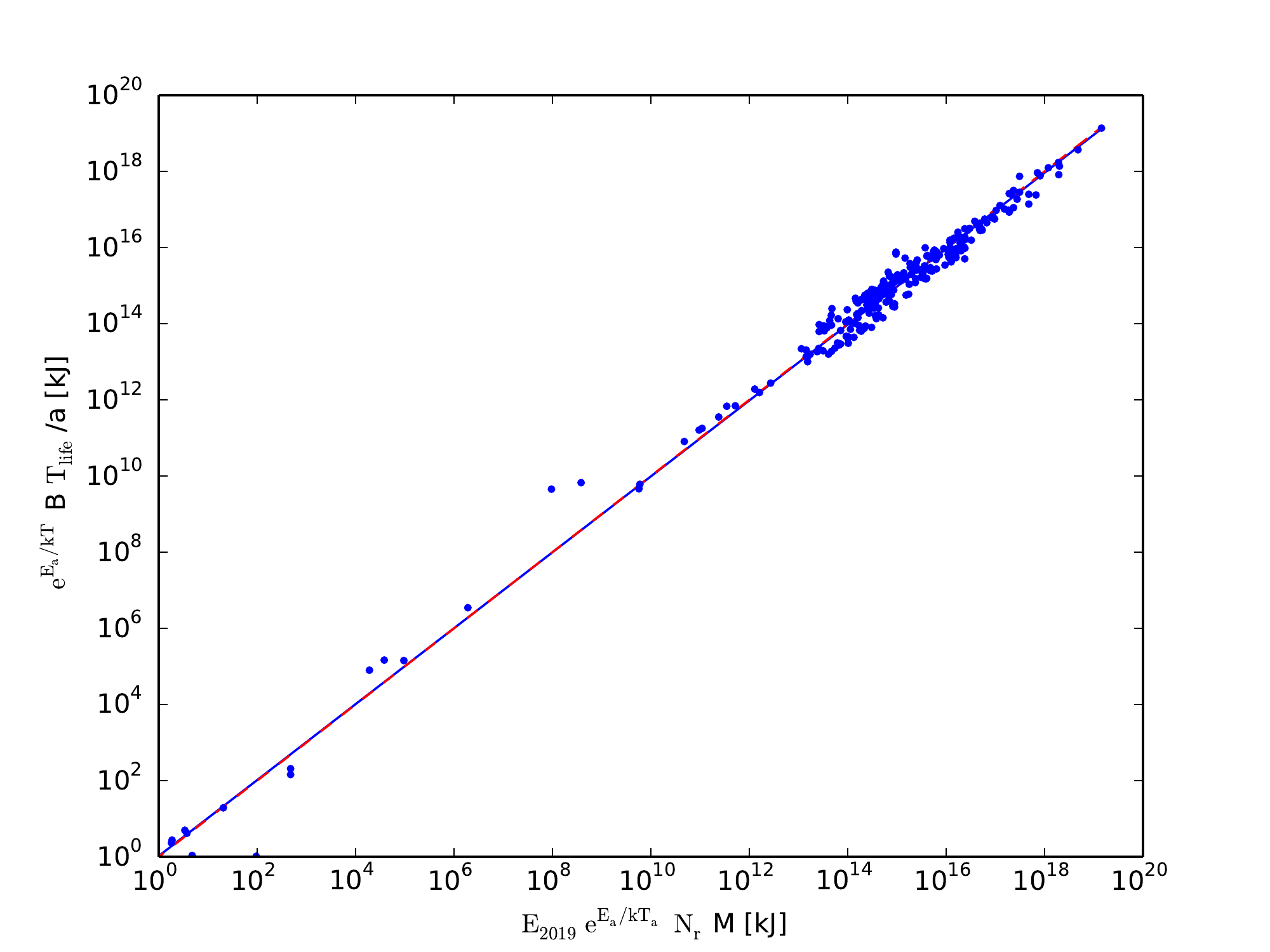}
\caption{Figure shows the  relation predicted by Eq. \ref{Eq1} for 277 living organisms, listed in ({\it 6}). The dashed red curve displays the identity given by  Eq \ref{Eq1}. The solid blue curve displays the best fitted value of slope 0.997 and normalization of 1.05 for ambient temperature  $\rm T_a = 30^o$C. The two curves are almost undistinguishable in the dynamical  range of 20 orders of magnitude.
}
\label{F1}
\end{center}
\end{figure}

The data displayed in Fig 1 strongly supports  the  unique  relation predicted by Eq. \ref{Eq1},  in a dynamical  range of 20 orders of magnitude, for all classes of organisms from $\it Bacteria$ to $\it Elephas \, Maximum$. The solid blue curve displays the best fitted value of slope 0.997 and normalization of 1.05, almost undistinguishable from the identity predicted by   Eq \ref{Eq1} (dashed red curve)
. The only free parameter (not predetermined by an independent measurement) in Eq \ref{Eq1} is the `ambient' temperature  $\rm T_a = 30^o$C, which was chosen only  to match the normalization in the best fitted  relation (solid blue) to the  identity 
(dashed red), 
but is a natural  choice for normalization, since ectotherms are typically around  $\rm 20^o$C and endotherms close to $\rm  40^o$C in Table 1 of  ({\it 6}). Moreover,  the slope close to unity (0.997) is independent of the $\rm T_a$ choice and for example, if we instead decide to preset $\rm T_a$  to the value of mammals ($\rm  37^o$C), it only changes the  best fitted normalization value  to  1.81. Therefore, the  relation between 5  physiological  variables ($\rm  t_{life}$, B, a, M \& T) given by  Eq \ref{Eq1} is confirmed without the choice of any free parameter. 

\begin{figure}[h!]
\begin{center}
\includegraphics[width=11.9cm]{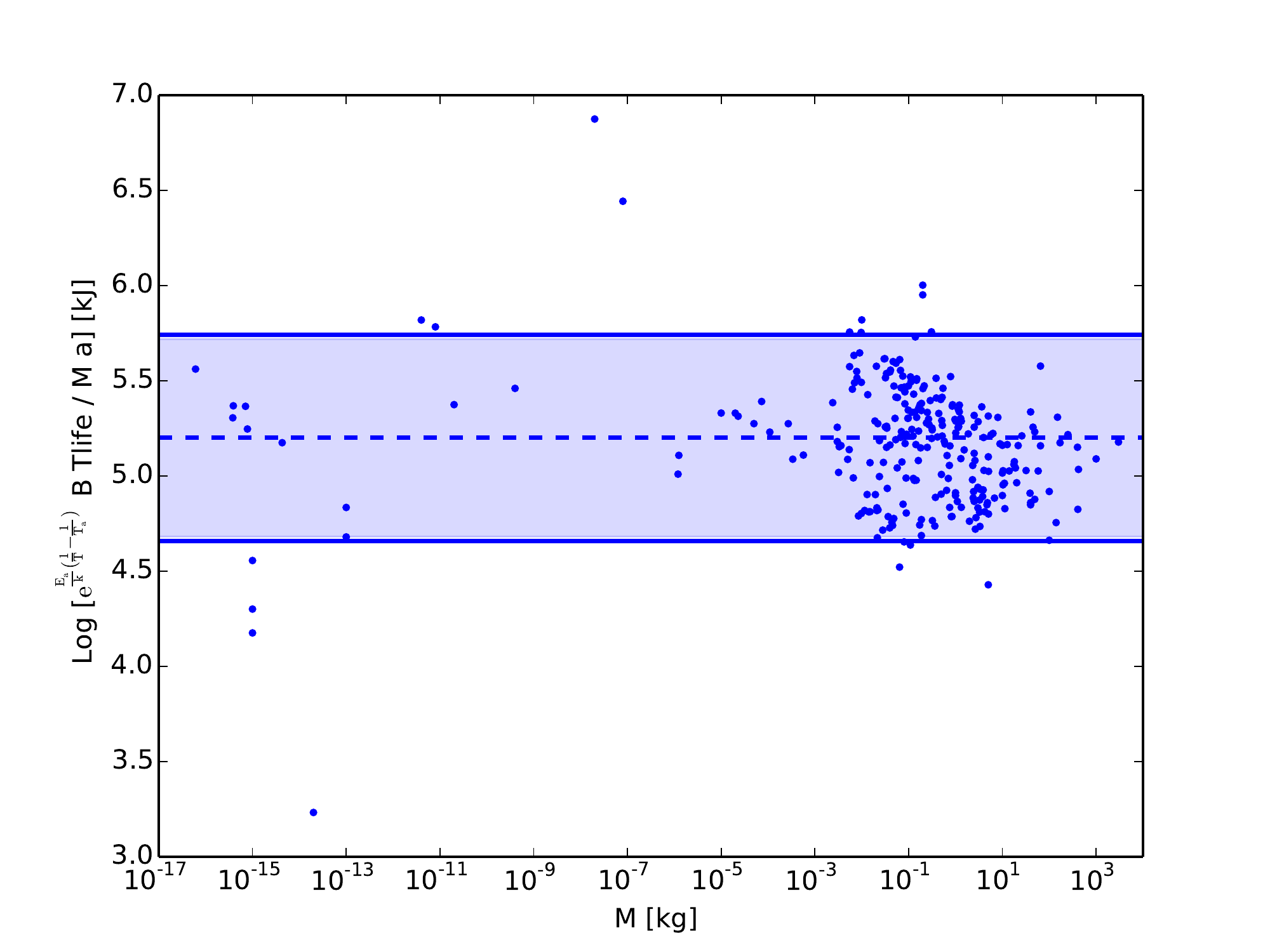}
\caption{Residuals  from the relation predicted by Eq \ref{Eq1} as a function of the organism's body mass for 277 living organisms, listed in  ({\it 6}).  The relation has only an average scatter of 0.339 dex  around the predicted value  ($\rm E_{2019} \, N_r $) denoted by the dashed line. The colored region between the solid curves denotes  residuals  less  than a geometrical factor of  $\pi$ from the relation and 95\% of the points ($\rm 2-\sigma$) fulfill such criterion.  
}
\label{F2}
\end{center}
\end{figure} 

Fig 2 shows the residuals  from the relation predicted by Eq \ref{Eq1}, as a function of the organism's body mass.  The relation has only an average scatter of 0.339 dex  around the predicted value  ($\rm E_{2019} \, N_r = E_{2019} \, N_b $/a;  dashed line in Fig 2), which impressively small taking into account  the 20 orders of magnitude variations  in  mass and that the values of $\rm E_{2019} =  10^{-3} \, Jg^{-1} $ ({\it 13}), $\rm N_b=7.3 \times 10^8$ heartbeats in a lifetime ({\it 16}) and a=4.5   ({\it 17}), comes from three completely independent  measurements in mammals. Moreover, around 95\% of the points ($\rm 2\,\sigma$) has departures from the relation less than a factor of  $\pi$ ($\approx \, \pm$0.5 dex, colored region between the solid curves in Fig 2). 

The extremely  accurate relation displayed in  Fig 2,  suggests that the only  missing  parameters are the ones  due to geometrical variations among species,  of the order of a dimensionless numerical factor of order unity ($\pi$). If that is indeed the case, this is probably the first relation in life sciences  including all  the relevant controlling parameters, reaching  an accuracy comparable to the ones in  exact sciences. Moreover, Fig 2  implies no clear trend (larger than an e-fold  $\approx \, $0.5 dex) in inter-class comparison from bacteria to the largest mammal. 

Inter-class variations was considered the most persuasive  evidence against the `rate of living'  theory, 
for example,  the inter-class comparison of birds and bats to mammals ({\it 2, 5}). These exceptions  were erased in Fig 2  due to  predicted  secondary parameters that are not included in the original formulation (mainly the parameter $\rm  a=   f_H/ f_{resp} $ for the particular case of birds and bats  to mammals).  Moreover, Eq \ref{Eq1} implies a  dependence of $\rm  t_{life} \propto \frac{M}{B} exp\Big({\small \frac{E_a}{k T_a}}\Big)$ that it is  in agreement with the three regimens experimentally known to extend life-span ({\it 20}): lowered ambient temperature  $\rm T_a $ in poikilotherms, decrease of physical activity in poikilotherms (lower B) and caloric restriction (lower B).

\section{Discussion}


We showed that the new metabolic rate relation  ({\it 13}) can be directly linked  to the total energy consumed in a lifespan, if it is conjectured a constant number $\rm N_r $ of respiration cycles per lifespan
, finding a corrected relation for the total energy consumed in a lifespan (Eq \ref{Eq1}) that  can  explain the origin of variations in the  `rate of living' theory  ({\it 2, 5}) and unify them into a single formulation. We  test   the validity and accuracy of the predicted relation  (Eq \ref{Eq1}) for the total energy consumed in a lifespan on $\sim$300 species representing all classes of living organisms, finding that the relation has an average scatter of only 0.3 dex, with 95\% of the organisms having departures less than a factor of $\pi$ from the relation, despite the $\sim$20 orders of magnitude difference in body mass. This reduces  any possible  inter-class variation in the relation to only a geometrical factor and strongly supports  the  conjectured  invariant  number of $\rm N_r \sim 10^8$ of respiration cycles per lifespan in all living organisms

Invariant quantities in physics traditionally   reflects fundamental underlying constraints, something   also applied recently to life sciences such as Ecology ({\it 21, 22}). Fig 2 displays the fact that, for a given temperature,   the  total life-span  energy consumption per gram per `generalized beat' ($\rm N_b^G \equiv a N_r = a  \,1.62 \times 10^8$) 
is remarkably  
constant on around $\rm E_{2019}$
, supporting that the overall  energetics during lifespan is the same for all living organisms
, being predetermined by the basic energetics of respiration. Therefore, Rubner's original picture  it is shown to valid  without systematic exceptions, but in this more general form. Moreover, since the  value determined from Fig 2 is remarkably similar to $\rm E_{2019}$, it can be considered an independent determination for $\rm E_{2019}$, suggesting that $\rm E_{2019}$  is a candidate for being an universal constant and not just a fitting parameter  coming from the corrected metabolic relation ({\it 13}).

In addition, 
we showed here that 
the  invariant total life-span  energy consumption per gram per `generalized beat'   comes directly from  the existence of another invariant: the approximately constant total number $\rm N_r \sim 10^8$ of respiration cycles per lifetime, effectively converting  the  `generalized beat' 
 into the characteristic clock during lifespan. Thus, the exact physical relation between (oxidative) free radical damage and the origin of aging,  is most probably related to the striking existence such of  constant  total number of respiration cycles $\rm N_r $ in the lifetime of all organisms, which predetermines the extension of life.  Therefore, since all organisms seems to live the same in units of respiration cycles,   future theoretical and experimental studies that investigate the exact link between the  constant  $\rm N_r \sim 10^8$   respiration cycles per lifespan and the production rates of free radicals (and other
byproducts of metabolism), should  shed light on the origin  of ageing and  the  
 physical   cause of natural mortality.
 

It have been also suggested that an analogous invariant  is originated at the molecular level ({\it 23}),  the number of ATP turnovers in a lifetime of the molecular respiratory complexes per cell, which from an energy conservation  model that extends metabolism to intracellular levels is estimated to be $\sim 1.5 \times 10^{16}$  ({\it 23}). Similar number can be determine taking into account that human cells requires synthase 
 approximately 100  moles of ATP daily, equivalent to $7 \times 10^{20}$  molecules per second.  For  $\sim 3 \times 10^{13}$ cells in human boby and for a respiration rate of 15 breaths per minute, this gives $\sim 9 \times 10^{7}$ ATP molecules synthesized per cell per breath, which for the invariant total number $\rm N_r $ of respiration cycles per lifetime found in this work, arises to the same number of  $\sim 1.5 \times 10^{16}$   ATP turnovers  in a lifetime per cell, showing the equivalence between both invariants, linking $\rm N_r $ to  the  energetics of respiratory complexes at cellular level.


The excellent  agreement between  the predicted relation (Eq 1) and the data across all types of living organisms      emphasizes  the fact that   lifespan indeed depends on multiple factors  (B, a, M, T \& $\rm T_a$) and strongly supports the  methodology presented in this work of multifactorial   testing, as done in Fig 1, since  quantities in life sciences generally suffers from a cofounding variable problem. An example of this problem, illustrated   on individually testing each of the relevant factors  is in ({\it 24}), which for a large (and noisy) sample test for $\rm  t_{life} \propto 1/B $, finding no clear correlation. From  Eq 1, it is clear that in an uncontrolled  experiment the dependence on the rest of the parameters  (M, a, T, $\rm  \,  \& \,  T_a$) might 
erase the dependence on the metabolic rate B  (in fact, for the same reason Rubner's work  ({\it 7}) focused on the mass-specific metabolic rate B/M instead of B). This work ({\it 24}) only finds a  residual inverse dependence of  $\rm  t_{life}$ on the ambient temperature $\rm T_a$ for ectotherms,  something expected   according to Eq 1 $\rm \Big( t_{life} \propto exp\Big({\small \frac{E_a}{k T_a}}\Big)\Big)$.

Finally, the empirical support in favor of   Eq \ref{Eq1} allow us to compute how much will vary the energy consumption in the biomass doing  aerobic respiration, as increases the earth's  temperature, relevant in the current context of possible global warming. This is  given by the factor  $\rm exp\Big[{\small \frac{E_a}{k} \Big(\frac{1}{ T}} - {\small \frac{1}{ T+1}}\Big) \Big]$ which for an activation energy  $\rm E_a = 0.63 eV$  and   temperature  of $\rm  30^o$C, implies an   increase 8.3\%  in energy consumption per 1 degree increase on the average Earth temperature. This  result can be straightforwardly  applied in  ectotherm since their body temperatures adapt to the environmental one ($\rm T=T_a$), but is less clear its implications for the case of endotherms organisms. 

\section*{Acknowledgments}

I acknowledge partial support from the Center of Excellence in Astrophysics and Associated Technologies (AFB-170002) and FONDECYT Regular Grant 1181663. 

\newpage

{\bf References and Notes}

\begin{enumerate}


\item Pearl, R. (1928). The Rate of Living. London: University of London Press.
\item Speakman, J.R. (2005). Body size, energy metabolism and lifespan. The Journal of Experimental Biology 208, 1717-1730.
\item   McCay, C. M.; Crowell, M. F.; Maynard, L. A. The effect of retarded growth upon the length of life span and upon the ultimate body size. J. Nutr. 10:63-79; 1935.
\item  Weindruch, R.; Walford, R. L. The retardation of aging and disease by dietary restriction. Springfield, IL: CC Thomas; 1988.
\item Ramsey, J. J., Harper, M. E. and Weindruch, R. (2000). Restriction of energy intake, energy expenditure, and aging. Free Rad. Biol. Med. 29, 946- 968.
\item Atanasov,  A. T.  (2012). Allometric Scaling Total Metabolic Energy per Lifespan in Living Organisms. Trakia Journal of Sciences, Vol. 10, No 3, pp 1-14, 2012
\item  Rubner, M. (1908). Das Problem det Lebensdaur und seiner beziehunger zum Wachstum und Ernarnhung. Munich: Oldenberg.
\item  Gerschmann, R., Gilbert, D. L., Nye, S. W., et al. (1954). Oxygen poisoning and x-irradiation: a mechanism in common. Science 119, 623-626.
\item Harman, D. (1956). Aging: a theory based on free radical and radiation biology. J. Gerontol. 11, 298-300.
\item Atanasov,  A. T.  (2007). The Linear Allometric Relationship Between Total Metabolic Energy per Life Span and Body Mass of Mammals, BioSystems, 90, 224-233.
\item Atanasov, A.T., 2005a. The linear allometric relationship between total metabolic energy per life span and body mass of poikilothermic animals. Biosystems 82, 137-142.
\item Atanasov, A.T., 2005b. Linear relationship between the total metabolic energy per life span and the body mass of Aves. Bulgarian Med. XIII, 30-32.
\item  Escala, A. (2019). The principle of similitude in biology. Theoretical Ecology 12 (4), 415-425.
\item Bridgman, P. W. (1922). {\it Dimensional Analysis}, Yale University Press.
\item Kleiber, M. (1932). Body size and metabolism, Hilgardia, 6, 315-351.
\item  Cook, S. et al. (2006). High heart rate: a cardiovascular risk factor? European Heart Journal, 27, 2387-2393.
\item Schmidt-Nielsen, K. (1984). {\it Scaling: why is animal size so important?},      Cambridge University Press.
\item   Brown, J. et al. (2004). Toward a Metabolic Theory of Ecology, Ecology, 85, 1771-1789.
\item Gillooly, J.F., Brown, J.H., West, G.B., Savage, V.M. \& Charnov, E.L. (2001). Effects of size and temperature on metabolic rate, Science, 293, 2248-2251.
\item  Sohal, R. S. and Weindruch, R. (1996). Oxidative stress, caloric restriction and aging. Science 273, 59-63.
\item Charnov, E. L. (1993). Life History Invariants: Some Explorations of Symmetry in Evolutionary Ecology, Oxford U. Press, New York 
\item Marquet, P.A. et al (2005). Scaling and power-laws in ecological systems, The Journal of Experimental Biology 208, 1749-1769
\item West, G. B., Woodruff, W. H. and  Brown, J. H. (2002). Allometric scaling of metabolic rate from molecules and mitochondria to cells and mammals, Proc Natl Acad Sci U S A., 99, 2473-2478
\item Stark, G., Pincheira-Donoso, D. and Meiri, S. (2020). No evidence for the `rate-of-living' theory across the tetrapod tree of life, Global Ecology and Biogeography, 00, 1-28

\end{enumerate}

\end{document}